\documentclass[11pt,american,twoside,a4wide]{article}
\usepackage[T1]{fontenc}
\usepackage[latin1]{inputenc}
\usepackage{latexsym}
\usepackage{amsmath}
\usepackage{bbm}
\usepackage{amssymb}
\usepackage{babel}
\usepackage{amsfonts}
\usepackage{times}
\usepackage{theorem}
\usepackage{epsfig}
\usepackage{enumerate}
\usepackage{color}
\usepackage[active]{srcltx}
\usepackage[colorlinks=true]{hyperref}
\usepackage{fancyhdr}
\newcounter{smallarabics}
\newenvironment{arabicenumerate}
{\begin{list}{{\normalfont\textrm{(\arabic{smallarabics})}}}
  {\usecounter{smallarabics}\setlength{\itemindent}{0cm}
   \setlength{\leftmargin}{5ex}\setlength{\labelwidth}{4ex}
   \setlength{\topsep}{0.75\parsep}\setlength{\partopsep}{0ex}
   \setlength{\itemsep}{0ex}}}
{\end{list}}

\newcounter{smallroman}

\newcommand{\ben}{\begin{arabicenumerate}}  
\newcommand{\een}{\end{arabicenumerate}}  

\newtheorem{theoreme}{Theorem}[section]
\newtheorem{proposition}[theoreme]{Proposition}
\newtheorem{lemma}[theoreme]{Lemma}
\newtheorem{definition}[theoreme]{Definition}
\newtheorem{corollary}[theoreme]{Corollary}

\def\rr{{\mathbb R}}
\def\zz{{\mathbb Z}}
\def\cc{{\mathbb C}}
\def\nn{{\mathbb N}}

\def\textsl{{}}

\def\Im{{\rm Im}\,}

\def\ch{{\frak h}}

\def\cT{\mathcal{T}}
\newcommand{\slim}{\mathop{\mathrm{s-lim}}\limits}

\def\c0inf{C_0^\infty}
\def\bep{\begin{proposition}}
\def\eep{\end{proposition}}

\def\proof{\noindent {\bf Proof.}\ \ }

\def\cH{{\cal  H}}

\def\cR{{\cal R}}

\def\i{{\rm i}}
\newcommand{\beq}{\begin{equation}}
\newcommand{\eeq}{\end{equation}}
\newcommand{\bear}[1]{\begin{array}{#1}}
\newcommand{\ear}{\end{array}}

\def\sp{{\hat e}}

\newcommand{\CAR}{\mathrm{CAR}}

\newcommand{\e}{\mathrm{e}}
\renewcommand{\i}{\mathrm{i}}

\renewcommand{\d}{\mathrm{d}}

\def\qed{$\Box$\medskip}

\def\cJ{{\cal J}}
\def\cT{{\cal T}}

\def\R{{\rm R}}

\def\bel{\begin{lemma}}
\def\eel{\end{lemma}}
\def\bet{\begin{theoreme}}
\def\eet{\end{theoreme}}
\def\bed{\begin{definition}}
\def\eed{\end{definition}}
\def\bar{\overline}
\def\ubar{\underline}

\def\12{\frac{1}{2}}

\def\x{\langle x \rangle}

\def\e{{\rm e}}

\def\d{{\rm d}}

\def\cH{{\cal H}}

\def\ac{{\rm ac}}

\def\sp{{\rm sp}}
\def\cS{{\cal S}}
\def\cR{{\cal R}}
\def\L{{l}}
\def\R{{r}}

\def\Ent{{\rm Ent}}

\def\fh{{\mathfrak h}}

\newcommand{\ds}{\displaystyle}
\begin{document}
\title{Landauer-B\"uttiker formula  and Schr\"odinger conjecture}
\author{L. Bruneau$^{1}$, V. Jak\v{s}i\'c$^{2}$, C.-A. Pillet$^3$
\\ \\ 
$^1$ D\'epartement de Math\'ematiques and UMR 8088\\
CNRS and Universit\'e de Cergy-Pontoise\\
95000 Cergy-Pontoise, France
\\ \\
$^2$Department of Mathematics and Statistics\\ 
McGill University\\
805 Sherbrooke Street West \\
Montreal,  QC,  H3A 2K6, Canada
\\ \\
$^3$Aix-Marseille Univ, CPT, 13288 Marseille cedex 9, France\\
CNRS, UMR 7332, 13288 Marseille cedex 9, France\\
Univ Sud Toulon Var, CPT, B.P. 20132, 83957 La Garde cedex, France\\
FRUMAM
}
\maketitle
\thispagestyle{empty}
%%%%%%%%%%%%%%%%%%%%%%%%%%%%%%%%%%%%%%%%%%%%%%%%%%
\begin{quote}
\noindent{\bf Abstract.}
We study the entropy flux in the stationary state of a finite one-dimensional sample $\cS$
connected at its left and right ends to two infinitely extended reservoirs $\cR_{l/r}$ at distinct 
(inverse) temperatures $\beta_{l/r}$ and chemical potentials $\mu_{l/r}$. The sample is a free 
lattice Fermi gas confined to a box $[0, L]$ with energy operator $h_{\cS,L }=-\Delta + v$. 
The Landauer-B\"uttiker formula expresses the steady state entropy flux in the coupled system 
$\cR_l+\cS+\cR_r$ in terms of scattering data. We study the behaviour of this steady state entropy 
flux in the limit $L\to\infty$ and relate persistence of transport to  norm bounds on the transfer 
matrices of the limiting half-line Schr\"odinger operator $h_\cS$. 
\end{quote}
%%%%%%%%%%%%%%%%%%%%%%%%%%%%%%%%%%%%%%%%%%%%%%%%%%
\section{Introduction}

This paper is part of the program initiated in \cite{AJPP2} and concerns transport in the so called
electronic black box model. This model describes a sample  $\cS$ (e.g., a quantum dot or
a more elaborate electronic device) coupled to several electronic reservoirs $\cR_j$. These reservoirs 
are free Fermi gas in thermal equilibrium at given temperatures and chemical potentials. In the 
independent electron approximation, the coupled system $\cS+\sum_j \cR_j$ is a free Fermi gas with 
single particle Hamiltonian $h=h_0+h_{\rm T}$, where $h_0$ is the single particle Hamiltonian of the 
decoupled system and $h_{\rm T}$ is the tunneling Hamiltonian describing the junctions coupling 
$\cS$ to the reservoirs. As time $t$ goes to infinity, the coupled system approaches a steady state 
which carries a non-trivial entropy flux. The celebrated Landauer-B\"uttiker formula gives a closed 
expression for this steady state entropy flux  in terms of the scattering data of the pair $(h,h_0)$. 
This formula was rigorously proven in the context of non-equilibrium quantum statistical mechanics 
relatively recently \cite{AJPP2, N}\footnote{We refer the reader to these papers for additional 
information on the Landauer-B\"uttiker formula and for references to the vast physics literature on 
the subject.}. Given the Landauer-B\"uttiker formula, the next natural question is the dependence of 
the steady state entropy flux on the structure of the sample $\cS$ (its geometry, its size, {\sl etc}). 
This paper is the first step in this direction of research. 

We consider the special case where $\cS$ is a finite one-dimensional structure described in the tight 
binding approximation by the single particle Hamiltonian $h_{\cS,L}=-\Delta_L+v$ on the Hilbert space
$\ell^2([0,L]\cap\zz)$. There $\Delta_L$ is the discrete Laplacian with Dirichlet boundary conditions 
and $v :\zz_+ \rightarrow \rr$ is a potential on the half line $\zz_+=\{0,1,\cdots\}$. This finite sample is 
coupled to two infinitely extended reservoirs, one at each of its boundary point. The resulting steady 
state entropy flux may vanish in the limit $L\to\infty$ and our goal is to characterize the persistence 
of transport in this limit in terms of the spectral data of the limiting half-line Schr\"odinger operator 
$h_\cS=-\Delta+v$ acting on  $\ell^2(\zz_+)$.

We start with a precise description of the model and the problem  we study. 
\pagestyle{fancy}

\subsection{Setup}
 
The electronic black box (EBB) model we consider in this paper is a special case of the class of 
models studied in \cite{AJPP2}, where the reader can find the proofs of the results described in 
this introductory section. A pedagogical introduction to the topic can be found in the  lecture 
notes \cite{AJPP1}.

Consider two free Fermi gases $\cR_l$ and $\cR_r$, colloquially called left and right reservoir, with 
single particle Hilbert space $\ch_l$ and $\ch_r$ and Hamiltonian $h_l$ and $h_r$. The single particle 
Hilbert space $\ch_\cS$ of the sample $\cS$ is finite dimensional and its single particle Hamiltonian is 
$h_\cS$. Until the very end of this section we shall  not need to further specify the structure of $\cS$. 
The EBB  model we shall study is a free Fermi gas with single particle Hilbert space
\[
\ch=\ch_l\oplus\ch_\cS\oplus\ch_r.
\]
The identity operators on $\ch$, $\ch_l$, $\ch_r$, $\ch_\cS$ will be denoted $1$,  $1_l$, $1_r$, 
$1_\cS$. Whenever the meaning is clear within the context, vectors and operators of the form 
$\psi\oplus 0$, $A\oplus 0$, \dots{} will be simply denoted by $\psi$, $A$,~\dots{} Accordingly,
$1_l$, $1_r$, $1_\cS$ will be identified with the corresponding orthogonal projections in $\ch$.
 
For $f\in\ch$, we denote by $a(f)/a^\ast(f)$ the annihilation/creation operators on the antisymmetric 
(fermionic) Fock space $\cH=\Gamma_-(\fh)$ over $\ch$. In the sequel, $a^\#(f)$ stands for $a(f)$ 
or $a^\ast(f)$. The Hamiltonian of the decoupled EBB system is $H_0=\d\Gamma(h_0)$, the second
quantization of
\[
h_0=h_l\oplus h_\cS\oplus h_r.
\]
The Hamiltonians and the number operators of the reservoirs are $H_{l/r}=\d\Gamma(h_{l/r})$ and 
$N_{l/r}=\d\Gamma(1_{l/r})$.

The algebra $\CAR(\fh)$ of canonical anticommutation relations over $\ch$ is the $C^\ast$-algebra
generated by the set of operators $\{ a^\#(f)\,|\, f \in \fh\}$. To any self-adjoint operator  $k$ on $\ch$ 
one associates the Bogoliubov group 
\[
{\rm b}_k^t(A)=\e^{\i t \d\Gamma(k)}A\e^{-\i t \d\Gamma(k)},
\]
of automorphisms of $\CAR(\fh)$. Note that 
\[
{\rm b}_k^t(a^\#(f))=\e^{\i t \d\Gamma(k)}a^\#(f)\e^{-\i t \d\Gamma(k)}=a^\#(\e^{\i t k}f).
\]
$\vartheta^t={\rm b}_1^t$ is the gauge group of the EBB model. We shall assume that the total 
charge $N=\d\Gamma(1)$ is conserved. The corresponding superselection rule distinguishes the 
gauge-invariant sub-algebra
\[
\CAR_{\vartheta}(\ch)=\{ A\in\CAR(\fh)\,|\, \vartheta^t(A)=A\text{ for all }t\},
\]
as the algebra of observables of the EBB model. The Bogoliubov group $\tau_0^t={\rm b}_{h_0}^t$ 
preserves $\CAR_{\vartheta}(\ch)$ and describes the time evolution of the decoupled EBB model. 
The pair $(\CAR_{\vartheta}(\ch), \tau_0^t)$ is a $C^\ast$-dynamical system.

For any self-adjoint operator $\varrho$ on $\fh$ satisfying $0\leq\varrho\leq1$ the formula 
\[
\omega_\varrho(a^\ast(f_n)\cdots a^\ast(f_1)a(g_1)\cdots a(g_n))=\det\{ \langle g_i,\varrho f_j\rangle\},
\]
defines a unique state $\omega_\varrho$ on $\CAR_\vartheta(\fh)$. It is called the quasi-free state 
of density $\varrho$ and is completely determined by its two point function 
\[
\omega_\varrho(a^\ast(f)a(g))=\langle g,\varrho f\rangle.
\]
The initial state of the EBB model  is the quasi-free state $\omega_0$ of density
\[
\varrho_l\oplus\varrho_{\cS}\oplus\varrho_r,
\]
where $\varrho_{l/r}$ denotes the Fermi-Dirac density at inverse temperature $\beta_{l/r}>0$ and 
chemical potential $\mu_{l/r}\in \rr$,
\beq
\varrho_{l/r}=\frac{1_{l/r}}{1_{l/r}+ \e^{\beta_{l/r}(h_{l/r}-\mu_{l/r}1_{l/r})}},
\label{initdensity}
\eeq
and $\varrho_\cS=1_\cS$ (none of our results depends on this particular choice of  $\varrho_\cS$). 
$\omega_0$ describes the thermodynamic state  in which the reservoirs $\cR_{l/r}$ are in thermal
equilibrium at inverse temperatures $\beta_{l/r}$ and chemical potentials $\mu_{l/r}$.

The coupling we will consider is specified by a choice of  non-zero vectors $\chi_{l/r}\in\ch_{l/r}$, 
$\psi_{l/r}\in \ch_\cS$. The left/right junction is described by the rank two operator 
\[
h_{T,\L/\R} = |\chi_{\L/\R}\rangle\langle\psi_{\L/\R}| +  |\psi_{\L/\R}\rangle\langle\chi_{\L/\R}|.
\]
The single particle Hamiltonian of the coupled EBB model is 
\[
h=h_0 + h_T=h_0+h_{T,\L} + h_{T,\R},
\]
and its Hamiltonian is
\[
H=\d\Gamma(h)= H_0+a^\ast(\psi_l)a(\chi_\L)+a^\ast(\chi_\L)a(\psi_l)
+a^\ast(\psi_r)a(\chi_r)+a^\ast(\chi_r)a(\psi_r).
\]
The dynamics of the coupled EBB model is described by the Bogoliubov group $\tau^t={\rm b}_h^t$. 
It preserves $\CAR_{\vartheta}(\ch)$ and the pair $(\CAR_\vartheta(\ch),\tau^t)$ is a 
$C^\ast$-dynamical system. The coupled EBB model is described by the quantum dynamical system
$(\CAR_\vartheta(\ch), \tau^t,\omega_0)$.

We now describe the energy/charge/entropy flux observables. Although the self-adjoint operators  
$H_{l/r}$ and $N_{l/r}$ are not in $\CAR(\ch)$, the differences 
\[
\Delta H_{l/r}(t)= \e^{\i t H}H_{l/r}\e^{-\i t H}- H_{l/r}, \qquad
\Delta N_{l/r}(t)= \e^{\i t H}N_{l/r}\e^{-\i t H}- N_{l/r},
\]
belong to $\CAR_{\vartheta}(\ch)$ for any $t\in\rr$, and one easily verifies the relations 
\[
\Delta H_{l/r}(t)=-\int_0^t \tau^s(\Phi_{l/r})\d s, \qquad 
\Delta N_{l/r}(t)=-\int_0^t\tau^s ({\cal J}_{l/r})\d s,
\]
where 
\begin{equation}
\label{eq:fluxobs}
\begin{split}
\Phi_{l/r}&=-\i[H,H_{l/r}]=\d\Gamma(-\i[h,h_{l/r}])=a^*(\i h_{l/r}\chi_{l/r})a(\psi_{l/r})
                                  +a^*(\psi_{l/r})a(\i h_{l/r}\chi_{l/r}),\\[3mm]
\cJ_{l/r}&=-\i [H,N_{l/r}]=\d\Gamma(-\i[h,1_{l/r}])=a^*(\i \chi_{l/r})a(\psi_{l/r})
                                  +a^*(\psi_{l/r})a(\i\chi_{l/r}).
\end{split}
\end{equation}

The self-adjoint operators $\Phi_{l/r}$, ${\cal J}_{l/r}$ belong to $\CAR_\vartheta(\ch)$ and are 
observables describing, respectively, the energy and charge flux out of the reservoir $\cR_{l/r}$. 
The associated entropy flux observable is 
\beq
\sigma=-\beta_l(\Phi_l -\mu_l {\cal J}_l)-\beta_r(\Phi_r-\mu_r{\cal J}_r).
\label{sigmadef}
\eeq
We recall the entropy balance equation \cite{JP, Ru}
\begin{equation}
\Ent(\omega_0\circ \tau^t|\omega_0)=-\int_0^t\omega_0(\tau^s(\sigma))\d s,
\label{ent-balance}
\end{equation}
where $\Ent(\,\cdot\,|\,\cdot\,)$ denotes Araki's relative entropy of two states \cite{Ar}%
\footnote{The entropy balance equation holds in a much wider context and is a very general 
structural property of non-equilibrium statistical mechanics.}. Since $\Ent(\,\cdot\,|\,\cdot\,)\leq 0$,
the balance equation ensures that  for all $t>0$ the average entropy flux is non-negative, 
\begin{equation}
\frac{1}{t}\int_0^t\omega_0(\tau^s(\sigma))\d s\geq 0,
\label{ent-balance1}
\end{equation}
in accordance with the second law of thermodynamics. 

A basic characteristic of out of equilibrium physical systems is the presence of non-vanishing steady 
energy, charge and entropy fluxes. Sharp mathematical results concerning the existence and values
of such fluxes can only be obtained in the idealization of the large time limit $t\rightarrow \infty$. 
To state the relevant result for the EBB model we need the assumption:

\begin{quote}
{\bf (H)} The single particle Hamiltonian $h$ has no singular continuous spectrum.
\end{quote}

\bet[\cite{AJPP2}]\label{ness}
Suppose that {\rm (H)} holds. Then for all $A\in \CAR_{\vartheta}(\ch)$ the 
limit 
\[
\omega_+(A)=\lim_{t\to\infty}\frac1t\int_0^t\omega_0(\tau^s(A))\d s,
\]
exists.
\eet
The functional $\omega_+$ is a state on $\CAR_{\vartheta}(\ch)$ and is called Non-Equilibrium 
Steady State (NESS) of the EBB model. The entropy balance equation (\ref{ent-balance1}) 
ensures  that $\omega_+(\sigma)\geq0$. The existence of $\omega_+$ is an 
open problem if $h$ has some singular continuous spectrum. 

Although the existence of a NESS for a given quantum dynamical system is generally a difficult 
analytical problem, the special quasi-free structure of the EBB model reduces the proof of 
Theorem \ref{ness} to  the study of the spectral and scattering theory of the pair $(h, h_0)$. 
Moreover, the steady state expectation values $\omega_+(\Phi_{l/r})$, $\omega_{+}({\cal J}_{l/r})$, 
$\omega_+(\sigma)$, can be expressed in closed form in terms of the scattering data of the pair 
$(h, h_0)$. The resulting expressions, the celebrated Landauer-B\"uttiker formulae, were rigorously 
proven in \cite{AJPP2,N} and yield natural necessary and sufficient conditions for the strict positivity
of $\omega_+(\sigma)$. We proceed to describe the Landauer-B\"uttiker formulae and the question 
we will study in this paper. 

We start with some basic observations about the EBB model. Let  $\tilde\ch_{l/r}\subset\ch_{l/r}$ be 
the cyclic subspace generated by $h_{l/r}$ and $\chi_{l/r}$ ({\sl i.e.,} the smallest $h_{l/r}$-invariant
subspace of $\ch_{l/r}$ containing $\chi_{l/r}$). The Hilbert space
\[
\tilde\ch=\tilde\ch_l\oplus\ch_\cS\oplus\tilde \ch_r,
\]
is invariant under $h$ and $h_0$, and $\Phi_{l/r},{\cal J}_{l/r},\sigma\in\CAR_{\vartheta}(\tilde\ch)$. 
Hence, for our purposes, w.l.o.g.\;we may replace $\ch_{l/r}$ and $\ch$ with $\tilde\ch_{l/r}$ and 
$\tilde\ch$ (we drop $\widetilde{\,\cdot\,}$ in the sequel). Let $\nu_{l/r}$ be the spectral measure for 
$h_{l/r}$ and $\chi_{l/r}$. By the spectral theorem we may assume that
$\ch_{l/r}=L^2(\rr, \d\nu_{l/r})$, $\chi_{l/r}(E)=1$ for all $E\in \rr$, and that  $h_{l/r}$ is the operator of
multiplication by the variable $E$. It follows that the density operator \eqref{initdensity} acts by 
multiplication with the function
\[
\varrho_{l/r}(E)=\frac{1}{1+\e^{\xi_{l/r}(E)}},\qquad \xi_{l/r}(E)=\beta_{l/r}(E-\mu_{l/r}).
\]
The absolutely continuous spectral subspace of $h_0$ is 
\[
\ch_{\ac}(h_0)=\ch_{\ac}(h_l)\oplus\ch_{\ac}(h_r)
=L^2(\rr, \d\nu_{l, \ac})\oplus L^2(\rr, \d\nu_{r, \ac}),
\]
where $\nu_{l/r,\ac}$ is the absolutely continuous part of $\nu_{l/r}$ (w.r.t.\;Lebesgue measure). 
To avoid discussion of trivialities we shall always assume that $\nu_{l/r,\ac}$ is non-zero (if either 
$h_l$ or $h_r$ has no absolutely continuous spectrum then
$\omega_+(\Phi_{l/r})=\omega_+({\cal J}_{l/r})=\omega_+(\sigma)=0$, see \cite{AJPP2}). 
The essential support  of the measure $\nu_{l/r, \ac}$, defined by,  
\[
\Sigma_{l/r}=\left\{E\in\rr\,\bigg|\,\frac{\d\nu_{l/r, \ac}}{\d E}(E)>0\right\},
\]
is also called the  essential support of the absolutely continuous spectrum of $h_{l/r}$. The intersection
of the supports
\[
\Sigma_{l \cap r}= \Sigma_l \cap \Sigma_r,
\] 
will play an important role in the sequel. As usual in measure theory, $\Sigma_{l/r}$ is only specified
up to a set of Lebesgue measure zero. More precisely, it is an equivalence class of the
relation 
$$
B_1\circeq B_2\Leftrightarrow|B_1\triangle B_2|=0,
$$
where $B_1, B_2$ are Borel sets and $|B|$ is the Lebesgue measure of $B$. As usual in measure 
theory we shall refer to such classes as sets.

Denote by $1_\ac(h_0)$ the orthogonal projection on $\ch_\ac(h_0)$. It follows from the trace class 
scattering theory that the wave operators
\[
w_{\pm}=\slim_{t\rightarrow \pm \infty}\e^{\i t h}\e^{-\i t h_0}1_\ac(h_0),
\]
exist. The scattering matrix $s=w_+^\ast w_- $ is a unitary on $\ch_\ac(h_0)$ and acts as the operator 
of multiplication by a unitary $2\times 2$ matrix function $s(E)$. We shall write this on-shell scattering 
matrix as
\[
s(E)=1+t(E)
\]
where
\[
t(E)=\left[\begin{array}{cc} t_{ll}(E) & t_{lr}(E) \\ t_{rl}(E) & t_{rr}(E)\end{array}\right],
\]
is the so-called $t$-matrix.
The entry $t_{lr/rl}(E)$ is the transmission amplitude from reservoir $\cR_{l/r}$ to the reservoir 
$\cR_{r/l}$ at energy $E$ and $|t_{lr/rl}(E)|^2$ is the corresponding transmission probability. We 
recall that, as a consequence of unitarity, $|t_{lr}(E)|^2=|t_{rl}(E)|^2$. We set $\cT(E)=|t_{lr}(E)|^2$
and notice that, as a consequence of formula (\ref{t-ampl})
\beq
\{ E\,|\,\cT(E)>0\}\circeq\Sigma_{l\cap r}.
\label{cTchar}
\eeq

\bet[\cite{AJPP2}] \label{lb-sick}
Suppose that {\rm (H)} holds. The steady state energy and charge currents are given by the following
Landauer-B\"uttiker formulae
\beq
 \omega_+( \Phi_{l/r})=\frac{1}{2\pi}\int_\rr \varphi_{l/r}(E)\d E,\qquad
 \omega_+(\cJ_{l/r})=\frac{1}{2\pi}\int_\rr j_{l/r}(E) \d E,
\label{eq:lb}
\eeq
where
\beq
 \varphi_{l/r}(E)=\cT(E) (\varrho_{l/r}(E)-\varrho_{r/l}(E))E,\qquad
 j_{l/r}(E)=\cT(E) (\varrho_{l/r}(E)-\varrho_{r/l}(E)).
\label{lb-fluxes}
\eeq
\eet
Thus, one can identify the functions $ \varphi_{l/r}$ and $ j_{l/r}$ as the
spectral densities of energy and charge current in the NESS $\omega_+$.
They satisfy the conservation laws 
\[
\varphi_l(E) +\varphi_r(E)=0, \qquad j_l(E) + j_r(E)=0.
\]
By Eq. \eqref{sigmadef}, the steady state entropy flux is given by
\beq
\omega_+(\sigma)=\frac{1}{2\pi}\int_\rr \varsigma(E)\d E,
\label{eq:lbsigma}
\eeq
where the spectral density
 \begin{equation}\label{lb-entropy}
 \begin{split}
 \varsigma(E)&= -\beta_l(\varphi_l(E)-\mu_l j_l(E)) -\beta_r(\varphi_r(E)- \mu_r j_r(E))\\[3mm]
 &=\cT(E)(\xi_r(E)-\xi_l(E))(\varrho_{l}(E)-\varrho_{r}(E)),
\end{split}
\end{equation}
is non-negative,  and
\[
\{ E\,|\,\varsigma(E)>0\}\circeq\{E\,|\,|\varphi_{l/r}(E)|>0\}\circeq\{E\,|\,|j_{l/r}(E)|> 0\}.
\]
If  $\beta_l=\beta_r$ and $\mu_l =\mu_r$ ({\em the equilibrium case}), then $\varphi_{l/r}$, $j_{l/r}$, 
and $\varsigma$ are zero functions. If either $\beta_l\not=\beta_r$ or $\mu_l \not=\mu_r$ 
({\em the non-equilibrium case}), then \eqref{cTchar} implies
\[ 
\{ E\,|\, \varsigma(E)>0\}\circeq \Sigma_{l \cap r}.
\] 
The functions $\varphi_{l/r}$, $j_{l/r}$ and $\varsigma$ are well defined and all the above properties 
hold even if $h$ has some singular continuous spectrum. However, the current state of the art results 
require Assumption (H) to link these functions to steady state currents and prove the 
Landauer-B\"uttiker formulae \eqref{eq:lb}.

Note that in the non-equilibrium case  $\omega_+(\sigma)>0$ iff $|\Sigma_{l \cap r}|>0$, {\sl i.e.,}
$\omega_+(\sigma)>0$ iff there exists an open scattering channel between $\cR_l$ and $\cR_r$.
Note also that even if $\omega_+(\sigma)>0$, it may happen that for some specific values of 
$\beta_{l/r}$, $\mu_{l/r}$ either $\omega_+(\Phi_{l/r})=0$ or  $\omega_+({\cal J}_{l/r})=0$. 
However, in the non-equilibrium case, $\omega_+(\Phi_{l/r})$ and  $\omega_+({\cal J}_{l/r})$ 
cannot simultaneously vanish and generically they are both different from zero.

We now describe the question we shall  study. Let $v: \zz_+\to\rr$ be a given potential. Consider the
finite lattice $\Gamma_L=[0, L]\cap\zz_+$ and suppose that the single particle Hilbert space and 
Hamiltonian of the sample  are $\ch_{\cS,L}=\ell^2(\Gamma_L)$ and $h_{\cS, L}=-\Delta_L + v_L$, 
where $(\Delta_Lu)(x)= u(x-1)+u(x+1)$ is the discrete Laplacian on $\Gamma_L$ with Dirichlet 
boundary conditions ({\sl i.e.,} $u(-1)=u(L+1)=0$) and $v_L$ is the restriction of the potential $v$ 
to $\Gamma_L$. The reservoirs $\cR_{l/r}$ and the vector $\chi_{l/r}$ are $L$ independent.
We take $\psi_l=\delta_0$, $\psi_r=\delta_L$ where $\delta_x$ denotes the usual Kronecker
delta at $x\in\Gamma_L$. We denote by $h_{T,L}$ the corresponding tunneling Hamiltonian
and set 
$$
h_L=h_{0,L}+h_{T,L},\qquad h_{0,L}=h_l\oplus h_{\cS,L}\oplus h_r.
$$
Denote by $\varphi_{l/r, L}$, $ j_{l/r, L}$ and $\varsigma_L$ the 
spectral densities of the steady state fluxes and let
\begin{equation}
\begin{split}
\bar{\frak T}&=\{ E\,|\, \limsup_{L\rightarrow \infty}\varsigma_L(E)>0\},\\[1mm]
\ubar{\frak T}&=\{ E\,|\, \liminf_{L\rightarrow \infty}\varsigma_L(E)>0\}.
\end{split}
\label{bas-def}
\end{equation}
Clearly, $\ubar{\frak T}\subset\bar{\frak T}\subset \Sigma_{l\cap r}$. Note also that 
\[
\begin{split}
\bar {\frak T}&=\{E\,|\,\limsup_{L\rightarrow \infty}|\varphi_{l/r, L}(E)|>0\}=
\{E\,|\,\limsup_{L\rightarrow \infty}|j_{l/r, L}(E)|>0\},
\end{split}
\]
and similarly for $\ubar {\frak T}$. 

Let $h_\cS=-\Delta +v$ be the limiting half-line Schr\"odinger operator acting on $\ell^2(\zz_+)$.
If $h_{\cS,L}$ is extended from $\ell^2(\Gamma_L)$ to $\ell^2(\zz_+)$ in the obvious way (by setting 
$h_{\cS,L}=0$ on $\ell^2(\Gamma_L)^{\perp}$), then $\lim_{L\to\infty}h_{\cS,L}=h_\cS$ in the strong 
resolvent sense. $\delta_0$ is a cyclic vector for $h_\cS$ and the corresponding spectral measure 
$\nu_\cS$ contains the full spectral information about $h_\cS$. The set 
\[
\Sigma_\cS=\left\{ E\,\bigg|\, \frac{\d\nu_{\cS,\ac}}{\d E}(E) >0\right\},
\]
is the essential support of the absolutely continuous spectrum of $h_\cS$. On  physical grounds, it is natural to introduce:
\begin{quote} 
{\bf Property RST.} The half-line Schr\"odinger operator $h_{\cS}$ exhibits regular spectral transport if for any choice of the reservoirs 
${\cal R}_{l/r}$, 
\begin{equation}
\ubar {\frak T}\circeq \bar {\frak T}\circeq \Sigma_{\cS}\cap \Sigma_{l\cap r}.
\label{conjecture1}
\end{equation}
\end{quote}
In the first version of this paper we have conjectured that Property RST holds for all potentials $v$ and we will 
comment further on this point in the next section.  If Property RST holds  and the reservoirs are 
chosen so that $\Sigma_\cS\subset\Sigma_{l\cap r}$, then $\Sigma_\cS$ is  precisely the set of 
energies at  which transport persists in the limit $L\to\infty$. Moreover, by Fatou's lemma, for any 
Borel set $B\subset\Sigma_\cS$ of positive Lebesgue measure, 
\[
\liminf_{L\to\infty}\int_B\varsigma_L(E) \d E >0,
\]
while the dominated convergence theorem implies 
\[
\lim_{L\to\infty}\int_{\rr\setminus\Sigma_\cS}\varsigma_L(E)\d E=0.
\]
Hence the essential support of the absolutely continuous spectrum of   operators satisfying (\ref{conjecture1}) has  a  physically  
natural  characterization in terms of transport.

Our main result gives sharp characterizations 
of the sets $\bar{\frak T}$ and $\ubar{\frak T}$ in terms of the growth of the norms of the transfer 
matrices associated to $h_\cS$. This characterization  shows that Property RST holds for  the potential $v$ if and only if  
the celebrated  Schr\"odinger conjecture (Property SC in the next section)  holds for $v$. This equivalence, which came as a surprise to us, links 
properties of generalized eigenfunctions   with the mechanism of non-equilibrium transport in this 
class of EBB models.
%%%%%%%%%%%%%%%%%%%%%%%%%%%%%%%%%%%%%%%%%%%%%%%%
\subsection{Results}

Since in the equilibrium case $\varsigma_L$ is  identically equal to zero, in what follows we assume 
the non-equilibrium case, {\sl i.e.,} that either $\beta_l\not=\beta_r$ or $\mu_l\not=\mu_r$. 

The transfer matrix at energy $E$ is defined by the product
\beq
T_L(E)=\left[\begin{array}{cc} v(L)-E & -1 \\ 1 & 0\end{array}\right]
\cdots\left[\begin{array}{cc} v(0)-E & -1 \\ 1 & 0\end{array}\right]. 
\label{Tmat}
\eeq
We denote by  ${\frak L}$  the collection of all sequences $(L_k)_{k\in\nn}$ of positive integers such that $L_k\uparrow \infty$.
Our main results is 
\bet\label{no-trampo}
There is a set $S$ in the equivalence class of $\Sigma_{l\cap r}$ such that, for any $E\in S$
and any $(L_k)_{k\in\nn}\in {\frak L}$, the following statements are equivalent.
\begin{enumerate}[{\rm (1)}]
\item $$\ds\lim_{k\to\infty}\varsigma_{L_k}(E)=0.$$
\item $$\ds\lim_{k\to\infty}\|T_{L_k}(E)\|=\infty.$$
\end{enumerate}
\eet
Let 
\[
{\frak S}_0=\left\{ E\,\bigg|\, \sup_{L}\|T_L(E)\|<\infty\right\},
\qquad 
{\frak S}_1=\left\{ E\,\bigg|\, \liminf_{L\to\infty}\|T_L(E)\|<\infty\right\}.
\] 

An immediate consequence of Theorem \ref{no-trampo} is 
\begin{corollary}\label{no-trampo-1}
\begin{enumerate}[{\rm (1)}]
\item
\[
\ubar {\frak T}\circeq {\frak S}_0 \cap \Sigma_{l\cap r}.
\]
\item
\[
\bar {\frak T}\circeq {\frak S}_1 \cap \Sigma_{l\cap r}.
\]
\item For any Borel set $B\subset{\frak S}_0\cap\Sigma_{l\cap r}$ of positive Lebesgue measure, 
\[
\liminf_{L\to\infty}\int_B\varsigma_L(E)\d E >0.
\]
\item
\[
\lim_{L\to\infty}\int _{\rr\setminus ({\frak S}_1\cap \Sigma_{l\cap r})}\varsigma_L(E)\d E =0.
\]
\end{enumerate}
\end{corollary}
 It follows from Corollary \ref{no-trampo-1} that  Property RST  is equivalent to 
\begin{quote} {\bf Property SC.} ${\frak S}_0 \circeq \Sigma_{\cS}\circeq {\frak S}_1$.
\end{quote}
Until recently, it was widely believed that Property SC holds for all potentials $v$ 
(see \cite{MMG} and Section C5 in \cite{S1}), a fact known as Schr\"odinger Conjecture. Regarding the existing results, the inclusion ${\frak S}_0 \subset \Sigma_{\cal S}$ was proven in 
\cite{GP, KP} (see also \cite{S2}). The inclusion $\Sigma_{\cal S}\subset {\frak S}_1$ was proven in \cite{LS}. 
After this work was completed  
and submitted for publication we have learned that Arthur Avila has announced a counterexample to the Schr\"odinger conjecture 
in the setting of  ergodic Schr\"odinger operators \cite{Av}.  

Property SC plays a central role in the spectral theory of one-dimensional Schr\"odinger operators.
Theorem \ref{no-trampo} and Corollary \ref{no-trampo-1} link this property, via 
the Landauer-B\"uttiker formula, to non-equilibrium transport and shed a new light on its 
physical interpretation.\footnote{We remark that to link Corollary \ref{no-trampo-1} with transport in non-equilibrium statistical mechanics
one needs that the Landauer-B\"uttiker formulae hold for all $L$ and hence that the coupled single 
particle Hamiltonian $h_L$ has no singular continuous spectrum for all $L$. A concrete example of reservoirs  where 
this is the case for any potential $v$  is $\ch_{l/r}=\ell^2(\zz_+)$, $h_{l/r}=-k\Delta$, $k>0$. For other examples and general results 
regarding this point we refer the reader to \cite{GJW}.}   Property SC appears very natural from the point of view of transport theory
and its failure provides examples of models with strikingly singular non-equilibrium
transport.  In particular, the transport properties of Avila's spectacular counterexample remain 
to be studied in the future.

\bigskip

{\noindent\bf Acknowledgment.}  
The research of L.B. and C.-A.P.  was partly supported by ANR (grant 09-BLAN-0098). The research 
of V.J. was partly supported by NSERC. A part of this work was done  during visits of the first and last 
authors to McGill University supported by NSERC and CNRS. Another part  was done during the stay of the
second author at University of Cergy-Pontoise. V.J. wishes to thank  V.~Georgescu and 
F.~Germinet for making this visit possible and for their hospitality. We wish to thank  A. Avila for making 
the manuscript \cite{Av} available to us and to Y. Last 
for useful discussions.

%%%%%%%%%%%%%%%%%%%%%%%%%%%%%%%%%%%%%%%%%%%%%%%%%%

\section{Proofs}
\label{sec:proof}
\subsection{Preliminaries}

We will denote by $\sp(A)$ the spectrum of a Hilbert space operator $A$, and write
$\Im A=(A-A^\ast)/2\i$. If $A$ is self-adjoint, then $\sp_\ac(A)$ denotes its absolutely
continuous spectrum and we write $A>0$ whenever $\sp(A)\subset]0,\infty[$.

In the following, we shall use indices $a,b,c,\ldots\in\{l,r\}$. We define
$$
F_a(z)=\langle\chi_a, (h_a- z)^{-1}\chi_a\rangle,
$$
and denote by $F(z)$ the $2\times2$ diagonal matrix with entries $F_{ab}(z)=\delta_{ab}F_a(z)$. 
We also introduce the $2\times2$ Green matrices $G_L^{(0)}(z)$ and $G_L(z)$ with entries
$$
G_{ab,L}^{(0)}(z)=\langle\psi_a, (h_{\cS,L}- z)^{-1}\psi_b\rangle,\qquad
G_{ab,L}(z)=\langle\psi_a, (h_L- z)^{-1}\psi_b\rangle.
$$

Next, we recall several basic facts regarding the boundary values of the resolvent and their role in
spectral theory. A pedagogical introduction to this topic, including complete proofs, can be 
found in \cite{J}. Let $A$ be a self-adjoint operator on a Hilbert space ${\frak H}$ and  
$\psi_1,\psi_2\in{\frak H}$. For Lebesgue a.e. $E\in \rr$
the  boundary values 
\begin{equation}
\langle \psi_1,(A-E-\i 0)^{-1}\psi_2\rangle=\lim_{\epsilon \downarrow 0} 
\langle \psi_1, (A-E-\i \epsilon)^{-1}\psi_2\rangle,
\label{boundary}
\end{equation}
exist and are finite. In the sequel, whenever we write $\langle \psi_1, (A-E-\i 0)^{-1}\psi_2\rangle$, 
we will always assume that the limit exists and is  finite. If the spectral measure $\nu_{\psi_1, \psi_2}$
for $A$ and $\psi_1$, $\psi_2$ is real-valued, then either $\psi_1$ is orthogonal to the cyclic subspace 
spanned by $A$ and $\psi_2$ and $\nu_{\psi_1, \psi_2}$ is the zero measure or
$\langle\psi_1,(A-E-\i 0)^{-1}\psi_2\rangle\not=0$ for Lebesgue a.e. $E\in\rr$.
If $\psi\in{\frak H}$ then $\Im\langle\psi,(A-E-\i 0)^{-1}\psi \rangle\ge0$ and if $\nu_\psi$ is the spectral
measure for $A$ and $\psi$, then
\[
\d\nu_{\psi, \ac}(E)=\frac{1}{\pi}\Im\langle\psi,(A-E-\i 0)^{-1}\psi \rangle\d E,
\]
so that the set $\{ E\,|\,\Im\langle\psi,(A-E-\i 0)^{-1}\psi\rangle>0\}$ is an essential support of 
$\nu_{\psi, \ac}$.

In particular, one has
\[
\d\nu_{l/r,\ac}(E)= \frac{1}{\pi}\Im F_{l/r}(E+\i 0)\d E,
\]
and, with a slight abuse of notation, we may denote the following concrete representative of the class 
$\Sigma_{l\cap r}$ by the same letter
$$
\{ E\,|\,\Im F(E+\i 0)>0\}=\Sigma_{l\cap r}.
$$
In words, $\Sigma_{l\cap r}$ consists of $E$'s  for which the boundary values $F_{l/r}(E+\i 0)$ exist, 
are finite, and have strictly positive imaginary part.

\subsection{Green's and transfer matrices}
It follows from stationary scattering theory (see \cite{Y}, Chap. 5) that  the $t$-matrix $t_L$ can be 
expressed in terms of the Green matrix $G_L$ by
\begin{equation}\label{t-ampl}
t_L(E)= 2\i(\Im F(E+\i 0))^{1/2}G_L(E+\i0)(\Im F(E+\i 0))^{1/2}.
\end{equation}
The formulae (\ref{t-ampl}) can be also proven directly by elementary means following the arguments 
in \cite{JKP}. The unitarity of the on shell scattering matrix $s_L(E)=1+t_L(E)$  implies that for 
Lebesgue a.e. $E\in \rr$, 
\beq
t_L^\ast(E)t_L(E)+t_L(E)+t_L^\ast(E)=0.
\label{optical}
\eeq
It follows that
\[
{\frak R}=\bigcap_L\{E\in\Sigma_{l \cap r}\,|\,\text{Eqs. \eqref{t-ampl} and \eqref{optical} hold}\},
\]
satisfies
\[
\Sigma_{l \cap r}\circeq {\frak R}.
\]
The following lemma relates the Green matrices $G_L^{(0)}$ and $G_L$.
\bel\label{dentist}
For $E\in\mathfrak{R}\setminus\sp(h_{\cS, L})$, one has 
$G_L^{(0)}(E)=(I-G_L^{(0)}(E)F(E+\i0))G_L(E+\i0)$.
\eel
\proof For $z\in\cc\setminus\rr$, the second resolvent formula
$$
(h_{L}-z)^{-1}-(h_{0,L}-z)^{-1}=-(h_{0,L}-z)^{-1}h_{T,L}(h_{L}-z)^{-1},
$$
yields
$$
G_{ab}(z)-G^{(0)}_{ab}(z)=-\sum_cG^{(0)}_{ac}(z)\langle\chi_c,(h_L-z)^{-1}\psi_b\rangle,
$$
and
$$
\langle\chi_c,(h_L-z)^{-1}\psi_b\rangle=-F_c(z)G_{cb}(z),
$$
which combine to give the desired formula.\hfill\qed

We proceed to relate the Green matrix $G_L^{(0)}$ with the transfer matrix \eqref{Tmat}.

\bel\label{TG0relat}
For $E\in\rr\setminus\sp(h_{\cS,L})$ and any $x,y,u,v\in\cc$ one has
$$
G_L^{(0)}(E)\left[\begin{array}{c}x\\y\end{array}\right]=\left[\begin{array}{c}u\\v\end{array}\right]
\Longleftrightarrow
T_L(E)\left[\begin{array}{c}u\\x\end{array}\right]=\left[\begin{array}{c}y\\v\end{array}\right].
$$
In other words, the permutation matrix $P^{(0)}:(x,y,u,v)\mapsto(u,x,y,v)$ maps the graph of 
$G_L^{(0)}(E)$ into that of $T_L(E)$.
\eel
\proof Fix $L$ and $E\in\rr\setminus\sp(h_{\cS,L})$. For $f\in\ell^2(\Gamma_L)$, the function
$\psi(x)=\langle\delta_x,(h_{\cS,L}-E)^{-1}f\rangle$ satisfies the finite difference equation
\beq
(-\Delta+v-E)\psi=f,
\label{fde}
\eeq
with boundary conditions $\psi(-1)=\psi(L+1)=0$. Using the transfer matrix
$$
T(x,y)=T_xT_{x-1}\cdots T_{y+1},\qquad 
T_j=\left[\begin{array}{cc} v(j)-E & -1 \\ 1 & 0\end{array}\right],
$$
the solution of the initial value problem for Equ. \eqref{fde} can be written as
$$
\left[\begin{array}{c}\psi(x+1)\\\psi(x)\end{array}\right]
=T(x,-1)\left[\begin{array}{c}\psi(0)\\\psi(-1)\end{array}\right]
-\sum_{z=0}^xT(x,z)\left[\begin{array}{c}f(z)\\0\end{array}\right].
$$
Setting $x=L$ and taking the boundary conditions into account yields
$$
T_L(E)\left[\begin{array}{c}\psi(0)\\0\end{array}\right]
-\left[\begin{array}{c}0\\\psi(L)\end{array}\right]
=\sum_{z=0}^LT(L,z)\left[\begin{array}{c}f(z)\\0\end{array}\right],
$$
which is an equation for the unknown $\psi(0)$ and $\psi(L)$. Setting $f=\delta_0$ and $f=\delta_L$,
we obtain the following equations for the entries of the matrix $G_L^{(0)}(E)$,
$$
T_L(E)\left[\begin{array}{c}G_{ll,L}^{(0)}(E)\\1\end{array}\right]
=\left[\begin{array}{c}0\\G_{rl,L}^{(0)}(E)\end{array}\right],
\qquad
T_L(E)\left[\begin{array}{c}G_{lr,L}^{(0)}(E)\\0\end{array}\right]
=\left[\begin{array}{c}1\\G_{rr,L}^{(0)}(E)\end{array}\right].
$$
Thus, the two linearly independent vectors $(G_{ll,L}^{(0)}(E),1,0,G_{rl,L}^{(0)}(E))$ and
$(G_{lr,L}^{(0)}(E),0,1,G_{rr,L}^{(0)}(E))$ span the graph of $T_L(E)$. One easily checks that
they are the images by the permutation matrix $P^{(0)}$ of the two vectors 
$(1,0,G_{ll,L}^{(0)}(E),G_{rl,L}^{(0)}(E))$ and
$(0,1,G_{lr,L}^{(0)}(E),G_{rr,L}^{(0)}(E))$ which span the graph of $G_L^{(0)}(E)$.
\hfill\qed

Combining the two previous lemmata, we obtain the connection between the transfer matrix and
the Green matrix $G(E+\i0)$.
\bel\label{TGrelat}
For $E\in\mathfrak{R}\setminus\sp(h_{\cS, L})$ and any $x,y,u,v\in\cc$ one has
$$
G_L(E+\i0)\left[\begin{array}{c}x\\y\end{array}\right]=\left[\begin{array}{c}u\\v\end{array}\right]
\Longleftrightarrow
T_L(E)\left[\begin{array}{c}u\\x+F_l(E+\i0)u\end{array}\right]
=\left[\begin{array}{c}y+F_r(E+\i0)v\\v\end{array}\right].
$$
In other words, the automorphism $P:(x,y,u,v)\mapsto(u,x+F_l(E+\i0)u,y+F_r(E+\i0)v,v)$ of $\cc^4$
maps the graph of $G_L(E+\i0)$ into that of $T_L(E)$.
\eel

%%%%%%%%%%%%%%%%%%%%%%%%%%%%%%%%%
\subsection{Proof of Theorem \ref{no-trampo}}

Formulas (\ref{lb-entropy}) and (\ref{t-ampl}) imply  that Theorem \ref{no-trampo} follows from
\bet\label{key}
Let $E \in {\frak R}\setminus(\cup_L\sp(h_{\cS, L}))\circeq \Sigma_{l \cap r}$
and $(L_k)_{k\in\nn}\in\mathfrak{L}$ be given. Then the following statements are 
equivalent.
\begin{enumerate}[{\rm (1)}]
\item 
\[
\lim_{k\to\infty}G_{lr,L_k}(E+\i0)=0.
\]
\item 
\[
\lim_{k \rightarrow \infty}\|T_{L_k}(E)\|=\infty.
\]
\end{enumerate}
\eet
\proof
(1) $\Rightarrow$ (2). We start with the observation that the unitarity relation \eqref{optical} implies
$\|t_L(E)\|\le2$. It follows from \eqref{t-ampl} that the sequence $\|G_{L_k}(E+\i0)\|$ is bounded.
Writing
$$
G_{L_k}(E+\i0)\left[\begin{array}{c}0\\1\end{array}\right]
=\left[\begin{array}{c}u_k\\v_k\end{array}\right],
$$
we conclude that the sequences $u_k$ and $v_k$ are bounded while (1) implies 
$u_k=G_{lr,L_k}(E+\i0)\to0$. It follows from Lemma \ref{TGrelat} that
$$
T_{L_k}(E)\left[\begin{array}{c}1\\F_l(E+\i0)\end{array}\right]
=\frac1{u_k}\left[\begin{array}{c}1+F_r(E+\i0)v_k\\v_k\end{array}\right],
$$
which clearly implies (2).

(2) $\Rightarrow$ (1). There exists bounded sequences $u_k$ and $x_k$ such that, writing
$$
T_{L_k}(E)\left[\begin{array}{c}u_k\\x_k+F_l(E+\i0)u_k\end{array}\right]
=\left[\begin{array}{c}y_k+F_r(E+\i0)v_k\\v_k\end{array}\right],
$$
the sequence $|v_k|+|y_k|$ diverges to infinity. By Lemma \ref{TGrelat}, one has
$$
G_{L_k}(E+\i0)\left[\begin{array}{c}x_k\\y_k\end{array}\right]
=\left[\begin{array}{c}u_k\\v_k\end{array}\right],
$$
and the boundedness of $\|G_{L_k}(E+\i0)\|$ implies that $|v_k|\le A+B|y_k|$ for some
positive constants $A$ and $B$. We conclude that $|y_k|\to\infty$ and (1) follows from
$$
G_{lr,L_k}(E+\i0)=\frac{u_k-G_{ll,L_k}(E+\i0)x_k}{y_k}.
$$
\hfill\qed
%%%%%%%%%%%%%%%%%%%%%%%%%%%%%%%%%%%%%%%%%%%%%%%%%

\end{document}